\documentclass{aa}
\usepackage{graphicx}
\usepackage{txfonts}
\usepackage{hyperref}
\usepackage[normalem]{ulem}
\hypersetup{colorlinks = true, citecolor = {blue}, urlcolor= {blue}}

\def\PGPU{$\varphi-$GPU }

\def\gapprox{\;\rlap{\lower 3.0pt                       
        \hbox{$\sim$}}\raise 2.5pt\hbox{$>$}\;}
\def\lapprox{\;\rlap{\lower 3.1pt                       
        \hbox{$\sim$}}\raise 2.7pt\hbox{$<$}\;}

\newcommand{\be}{ \begin{equation} }
\newcommand{\ee}{\end{equation}}

\newcommand{\ben}{\begin{enumerate}}
\newcommand{\een}{\end{enumerate}}

\renewenvironment{thebibliography}[1]{\begin{oldthebibliography}{#1}\setlength{\parskip}{0ex}\setlength{\itemsep}{0ex}}{\end{oldthebibliography}}

\usepackage{diagbox}
\usepackage{siunitx}
\newcommand{\orcid}[1]{\href{https://orcid.org/#1}{\protect\includegraphics[width=8pt]{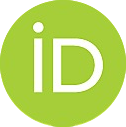}}}
\makeatletter
\renewcommand*\aa@pageof{, page \thepage{} of \pageref*{LastPage}}
\makeatother

\def\ba{\mbox{\boldmath $a$}}
\def\acck#1{\ba_i^{(#1)}}
\def\absak#1{|\acck{#1}|}

\usepackage{amstext}

\begin{document}
   
\title{Milky Way globular clusters on cosmological timescales. III. Interaction rates}

\author{Maryna~Ishchenko
\inst{1,2,4}\orcid{0000-0002-6961-8170 }
\and
Margaryta~Sobolenko
\inst{1,4}\orcid{0000-0003-0553-7301}
\and
Peter~Berczik
\inst{3,4,5,1}\orcid{0000-0003-4176-152X}
\and
Chingis~Omarov
\inst{2}\orcid{0000-0002-1672-894X}
\and
Olexander Sobodar
\inst{1,4}\orcid{0000-0001-5788-9996}
\and
Mukhagali Kalambay
\inst{2,6,7}\orcid{0000-0002-0570-7270}
\and
Denis~Yurin
\inst{2}\orcid{0000-0002-5604-9757}
}

\institute{Main Astronomical Observatory, National  
           Academy of Sciences of Ukraine,
           27 Akademika Zabolotnoho St, 03143 Kyiv, Ukraine
           \email{\href{mailto:marina@mao.kiev.ua}{marina@mao.kiev.ua}}
           \and
           Fesenkov Astrophysical Institute, 050020, Almaty, Kazakhstan
           \and
           Astronomisches Rechen-Institut, Zentrum f\"ur Astronomie, University of Heidelberg, M\"onchhofstrasse 12-14, 69120 Heidelberg, Germany
           \and
           Nicolaus Copernicus Astronomical Centre, Polish Academy of Sciences, ul. Bartycka 18, 00-716 Warsaw, Poland
           \and 
           Konkoly Observatory, Research Centre for Astronomy and Earth Sciences, E\"otv\"os Lor\'and Research Network (ELKH), MTA Centre of Excellence, Konkoly Thege Mikl\'os \'ut 15-17, 1121 Budapest, Hungary
           \and
           Al-Farabi Kazakh National University,
           71 al-Farabi ave., 050040 Almaty, Kazakhstan
           \and 
           Energetic Cosmos Laboratory, Nazarbayev University,
           53 Kabanbay Batyr ave., 010000 Astana, Kazakhstan
           }

\date{Received xxx / Accepted xxx}
    
\abstract    
{}
{We carry out the self-consistent dynamic evolution of the orbital structure of Milky~Way globular clusters. This allows us to estimate possible and probable close passages and even collisions of the clusters with each other.}
{We reproduced the orbits of 147~globular clusters in 10~Gyr lookback time using our own high-order $N$-body parallel dynamic $\varphi$-GPU code. The initial conditions (three coordinates and three velocities for the present time) were derived from the \textit{Gaia}~DR3 catalogue. For each of the GCs, 1000 initial conditions were additionally generated, taking the \textit{Gaia} measurement errors into account. The  galaxy is represented by five external potentials from the IllustrisTNG-100, whose masses and sizes of the disk and halo components are similar to the physical values of the Milky~Way at present.}
{We present a statistical analysis of the cumulative close passages rate: $\text{About ten}$ close passages with relative distances shorter than 50 pc for every billion years for each of the five external potentials. We present the 22 most reliable collision pairs with a good probability. As an example: Terzan~4 versus Terzan 2 (49\%), Terzan~4 versus NGC~6624 (44\%), Terzan~4 versus Terzan~5 (40\%), Terzan~4 versus NGC 6440 (40\%), and Terzan~4 versus Liller~1 (42\%). The most active globular cluster in the collision sense is Terzan 4, which has 5.65 collision events on average (averaged over all individual 1000 initial condition realisations). Most collisions are located inside the Galactic disk and form two ring-like structures. The first ring-like structure has the highest collision number density at $\sim$1~kpc, and the second sturcture has a maximum at $\sim$2~kpc.}
{Based on our numerical simulations, we can conclude that the few dozen Milky~Way globular clusters probably undergo some close encounters and even possible collisions during their lifetimes, which can significantly affect their individual dynamical evolution and possibly even their stellar content.}


\titlerunning{Interaction rates of GCs}
\authorrunning{M.~Ishchenko et al.}
\maketitle

\section{Introduction}\label{sec:Intr}

According to the $\Lambda$CDM model, the Milky Way (MW) globular clusters~(GCs) are one of the first bound stellar systems formed in the early Universe as gravitationally bound stellar systems \citep{Weinberg1993} with typical ages of about 10-12~Gyr \citep{2009ApJ...694.1498M,VandenBerg2013, Valcin2020} and current masses $\gtrsim10^{5}\rm M_{\odot}$ \citep{Harris2013,Kharchenko2013, Baumgardt2019, Baumgardt2021}. About 160 GCs are found in the MW today, whose parameters based on \textit{Gaia} data~\citep{Gaia2021} are measured with a relatively good precision, including the full 6D phase-space information~\citep{VasBaum2021, Baumgardt2021}.  

During the orbital analysis of GCs in our previous papers (\cite{Ishchenko2023a} (hereafter \hyperlink{I23}{\color{blue}{Paper~I}}) and \cite{Ishchenko2023c}), we  observed that some GCs have a large number of orbital revelations over 10 Gyr, for example, NGC 6624 has $\sim$450 turns, and Terzan 4 has $\sim$620, and some of them are associated with the specific Galaxy regions (bulge or disk). 
Due to the long-term presence of these objects in same Galactic space volume (including even the Galactic center; \citealt{Ishchenko2023c}), close interactions of these objects with each other are expected. Based on these general considerations, the probability of a close meeting of our GCs cannot be neglected. 

In our preliminary investigations based on Gaia DR2, we analysed the possible orbital crossing or collision of 150 GCs. \cite{Chemerynska2022} integrated the GC orbits up to 5 Gyr in the fixed MW-like gravitational potential, that is, in this case the masses, scale lengths, and scale heights of the MW external potential were not evolved. In these simple conditions, we found several GC collision pairs with a probability higher than 20\%. \cite{Ishchenko2023-kfnt} integrated the GCs orbits up to 10 Gyr in an MW-like gravitational potential that changed in time, that is, the masses, scale lengths, and scale heights were evolved. The values of the measurement errors from \textit{Gaia} DR2 for radial velocities and proper motions were also taken into account to obtain statistically more significant result. We also found several GC collision pairs that probably interact closely with each other. 

A similar type of investigation was carried out by \cite{Khoperskov2018}, who analysed the close passages of thick-disk GCs and presented the collision rate. \cite{Marcos2014} also  analysed the possible influence of GC close passages on the formation and/or distraction of open clusters. The authors presented GC FSR 1767 (2MASS-GC04) as a candidate destroyer of the open cluster on a long timescale.  

Although the MW GC subsystem is usually assumed to be collisionless, more recent studies showed that many of them have disk-like kinematics~\citep{Casetti-Dinescu2010, VandenBerg2013}, which enables their close orbital passages and gravitational interactions with possible rotation angular momentum gain. These interactions may result in a mutual mass-exchange and in changes in orbital motions~\citep{Khoperskov2018}. Another source of an orbital transformation of GCs is the mass growth of the Milky Way, whose stellar content increased by a factor of $10$ in the last $10$~Gyr~\citep{Bajkova2021AstL, Garrow2020, Boldrini2022}, where the increase in mass should squeeze the orbits of the GCs over time and might increase the probability of close passages or collisions of the GCs. Finally, massive satellites orbiting the MW can impact the gravitational field of the Galaxy and also affect the motion of a stellar system in the halo and the disk~\citep{Garrow2020, Boldrini2022}. Finally, if the GCs were more massive at earlier epochs, their gravitational interaction might lead to the scattering of their orbits, which is especially important in the innermost parts of the MW disk or in the bulge region.

The paper is organised as follows. In Section~\ref{sec:integr} we introduce the initial conditions of GC data sampling with an overview of the MW-like potentials and integration procedure. In Section~\ref{sec:gc-timescale} we  estimate the GC collision probability with different selection criteria. We also present the distribution of GC collisions by Morton ordering and in a 3D Cartesian galactocentric coordinate system. In Section~\ref{sec:con} we present the results of our selected GC sample and summarise our findings.

\section{Initial conditions and integration procedure}\label{sec:integr}

Before the orbital integration, we analysed the GC \textit{Gaia} measurement errors from the catalogues of \cite{VasBaum2021} and \cite{Baumgardt2021} , which contain information about more than 160~objects. The catalogues contain the complete 6D phase-space information required for the initial conditions of our simulations: right ascension~(RA), declination~(DEC), heliocentric distane~(D$_{\odot}$), proper motion in right ascension (PMRA), proper motion in declination (PMDEC), and the radial velocity (RV). After the analysis of the error measurements for PMRA, PMDEC, RV, and D$_{\odot}$ , we excluded GCs with large uncertainties (relative errors more that 30\%)~(see \cite{Baumgardt2021}\footnote{\label{note1}The error values for PMRA, PMDEC, RV, and D$_{\odot}$ are taken from \url{https://people.smp.uq.edu.au/HolgerBaumgardt/globular/orbits_table.txt}}: Pal~1, AM~1, NGC~2419, Pal~3, Crater, AM~4, NGC~6380, NGC~6553, BH~261, NGC~6760, Lae~3, and Pal~13. We also excluded the GC Mercer 5 because no mass information for this GC is available in the catalogues above: During the dynamical simulation of the GCs, we used the clusters with the self-gravity together with the Galaxy external potential. Thus, we finally obtained a sample of N$_{GC}$ = 147 GCs for the future integration and analysis. 

Although we excluded objects with large errors from further analysis, small errors in the measurements can lead to a change in the shape of the orbit if the orbital integration is performed over a long period of time. Thus, in order to obtain statistically significant results, we performed 1000 simulations in which we varied the initial velocities (PMRA, PMDEC, and RV) and heliocentric distance (D$_{\odot}$) of the GCs using the normal distribution of the corresponding measurement errors ($\pm\sigma$) taken from the catalogue above. We carried out 5000 simulations: 1000 times  five TNG-TVPs. 

\begin{figure*}[htbp]
\centering
\includegraphics[width=0.95\linewidth]{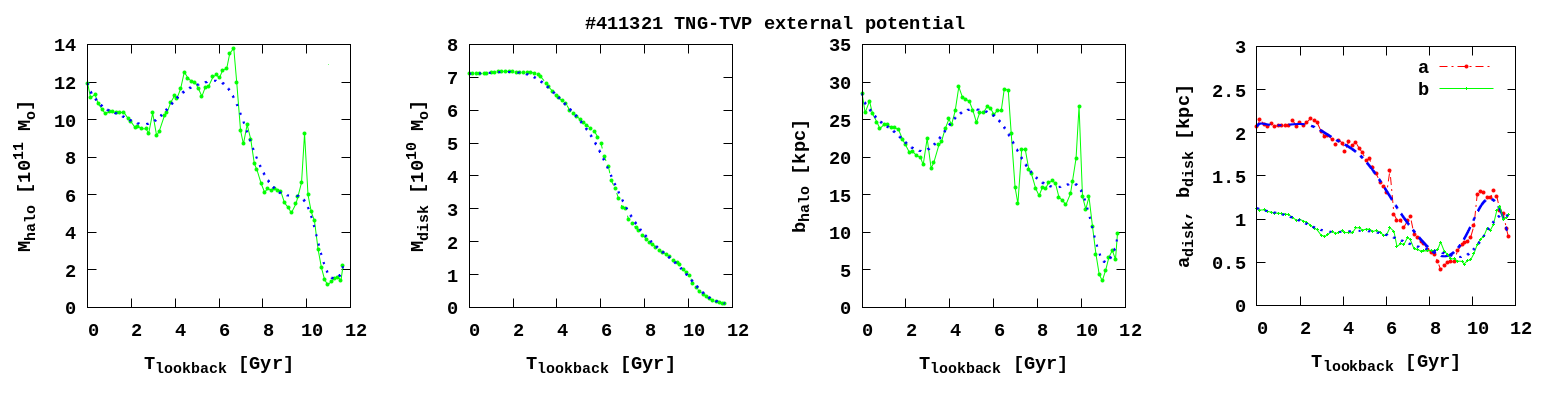}
\caption{Evolution of halo and disk masses, and their characteristic scales for {\tt 411321} TNG-TVP as a typical example. \textit{From left to right:} Halo mass $M_{\rm h}$, disk mass $M_{\rm d}$, Navarro-Frenk-White halo scale parameter $b_{\rm h}$, and Miyamoto-Nagai disk-scale parameters $a_{\rm d}$ and $b_{\rm d}$. The solid green lines with dots show the parameters recovered from the IllustrisTNG-100 data. The dotted and dash-dotted blue lines correspond to the values after the interpolation and smoothing with a 1~Myr time-step that was used in the orbital integration.}
\label{fig:MW-TNG}
\end{figure*}

To perform a more realistic integration of our MW GC orbits, we used the publicly available data from the IllustrisTNG-100 cosmological magnetohydrodynamical simulations~(\cite{2018MNRAS.473.4077P,2018MNRAS.475..624N,2018MNRAS.475..676S,2018MNRAS.480.5113M,2018MNRAS.477.1206N,2019ComAC...6....2N}). From the simulated MW-like potentials in the IllustrisTNG-100, we selected five galaxies that reproduce the currently best Milky Way parameters (\citealt{Mardini2020}, \hyperlink{I23}{\color{blue}{Paper~I}}). From the simulated IllustrisTNG-100 galaxies, we selected our five objects ({\tt 411321, 441327, 451323, 462077} and {\tt 474170}) that reproduce the MW basic parameters well at zero redshift~(disk and halo masses with the spatial scales). In Fig.~\ref{fig:MW-TNG} we present the evolution of halo and disk masses as an example and their characteristic scales for external potential of {\tt 411321} selected from the IllustrisTNG-100 cosmological database. The evolution of another four TNG-100 time-variable potentials (TNG-TVP) are presented on the web-page \footnote{MW-like potentials are taken from \url{https://bit.ly/3b0lafw}} , and details about the selection procedure and fitting can be found in \hyperlink{I23}{\color{blue}{Paper~I}}. 

For the GC orbital integration, we used the high-order parallel $N$-body code $\varphi$-GPU \footnote{$N$-body code \PGPU: \\~\url{ https://github.com/berczik/phi-GPU-mole}} , which is based on the fourth-order Hermite integration scheme with hierarchical individual block time-steps \citep{Berczik2011,BSW2013}. 
The external gravity was also added to the code (see more details in our previous publication (Section 2.2 and 2.3, \hyperlink{I23}{\color{blue}{Paper~I}})). 

The present-day positions and velocities of the 147 MW GCs were used as the initial conditions for the backward integration for $10$ Gyr. The procedure of transformation from the DR3 reference frame to the Cartesian galactocentric rest-frame for GCs initial positions and velocities is also described in \cite{Chemerynska2022} and \hyperlink{I23}{\color{blue}{Paper~I}}. 

For the transformation of the positions and velocities into the Cartesian galactocentric rest-frame (see \citealt{Johnson1987} and \citealt{Bovy2011} for the coordinate transformation equations), except for the equatorial position of the North Galactic Pole (NGP), we used the updated values RA$_{\rm NGP}=192\fdg7278$, DEC$_{\rm NGP}=26\fdg8630$, and $\theta_{\rm NGP}=122\fdg9280$ \citep{Karim2017}. We assumed a galactocentric distance of the Sun of $R_{\odot}=8.178$~kpc \citep{Gravity2019, Reid2004}, a height above the Galactic plane of $Z_{\odot}=20.8$~pc \citep{Bennett2019}, and a velocity of the local standard of rest (LSR) of $V_{\rm LSR}=234.737$~km~s$^{-1}$ \citep{Bovy2012, Drimmel2018}. Accordingly, the Sun is located at $X_{\odot} = -8178$~pc, $Y_{\odot} = 0$~pc and $Z_{\odot} = 20.8$~pc in our Cartesian galactocentric coordinate system. For the peculiar velocity of the Sun, we used the following values with respect to the LSR: $U_{\odot}=11.1$~km~s$^{-1}$, $V_{\odot}=12.24$~km~s$^{-1}$, and $W_{\odot}=7.25$~km~s$^{-1}$~\citep{Schonrich2010}.

The detection of the GC close passages can be sensitive to the integration time-step parameter $\eta$ \citep{MA1992}. In the particular case of the fourth-order Hermite integration scheme, a time-step can be written as
\begin{eqnarray}
\label{eq:aarseth-timestep}
\Delta t_{i} &=& \sqrt{\eta} \cdot \frac{A_i^{(1)}}{A_i^{(2)}}, \\
A_i^{(k)} &=& \sqrt{\absak{k-1}\absak{k+1} + \absak{k}^2},
\end{eqnarray}
where $\boldsymbol{a}^{(k)}$ is the $kth^{\rm }$ derivative of the $i$th particle acceleration. Thus, the time-steps are directly proportional to the $\eta$ parameter, which is finally is responsible for the total integration accuracy itself. For the highest-order integration Hermite scheme, the generalised Aarseth criterion can be found in \cite{Nitadori2008}.  

\begin{figure}[htbp!]
\centering
\includegraphics[width=0.9\linewidth]{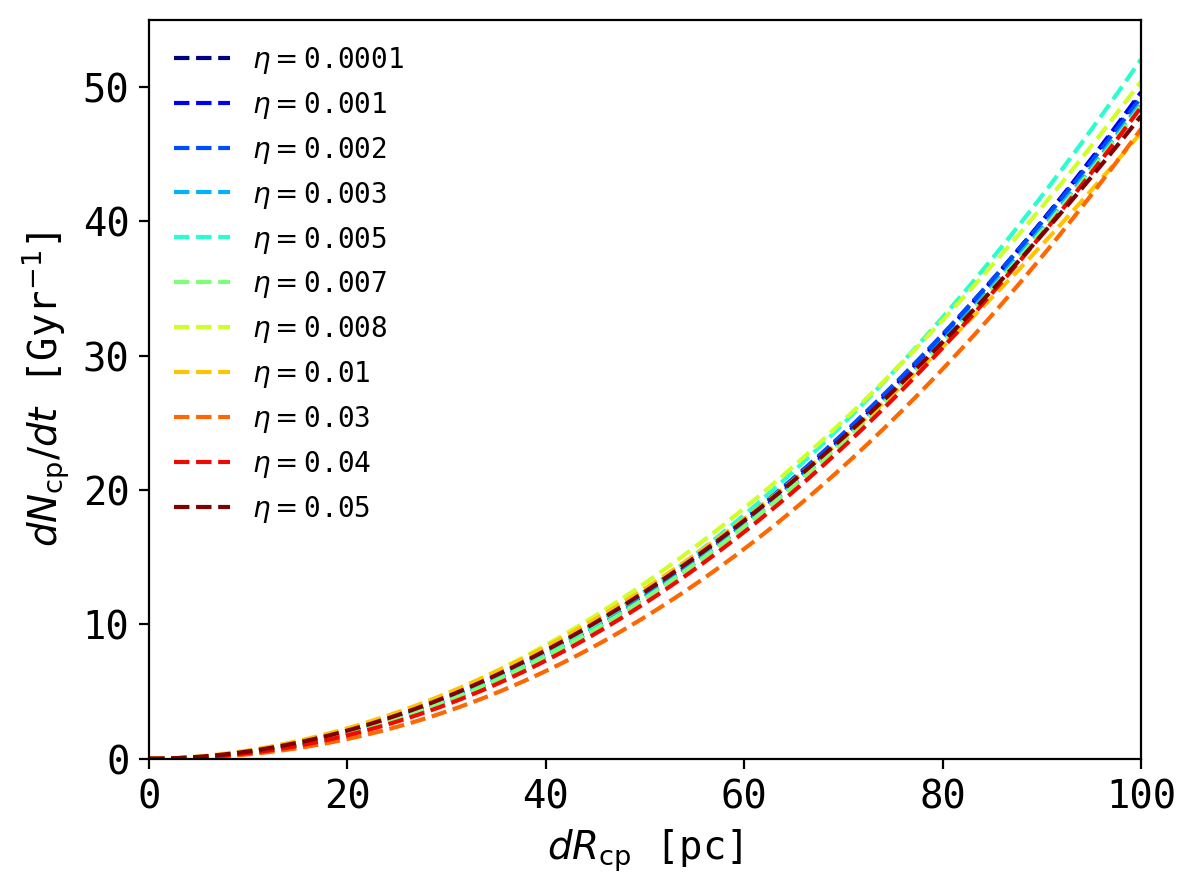}
\caption{Comparison of the GC global close passage rate in the same initial configuration, but with $11$ different timestep parameters $\eta$ in the {\tt 411321} TNG-TVP external potential.}
\label{fig:eta}
\end{figure}

Using our initial conditions, we made additional calculations for $11$ different $\eta$ parameters: 0.01, 0.03, 0.04, 0.05, 0.001, 0.002, 0.003, 0.005, 0.007, 0.008, and 0.0001. For lower values of $\eta$, the integration time-step is proportionally smaller, which accordingly means more individual points on each GC orbit. For this set of runs, we used the same initial position and velocities for all additional $11$ calculations in the selected {\tt 411321} external potential. The resulting global close passage rates are presented in the Fig.~\ref{fig:eta} (for details of the close passage, see Section~\ref{subsec:cp}).

 Fig.~\ref{fig:eta} shows that the integration parameter accuracy value $\eta$ affects the close passage rate very little.  $\eta=0.01$ limits the total relative energy drift ($\Delta E_{\rm tot}/ E_{{\rm tot}, ~t=0}$) over a 10~Gyr backward integration to below $\approx2.5\times10^{-13}$. The typical integration time (on a desktop system AMD Ryzen Threadripper 3960X 24 core with ten parallel threads) for $\eta$=0.01 takes approximately 21~minutes instead of 1080~minutes for $\eta$=0.0001 for a single run. The wallclock time calculation for the main set of 1000 production runs in one of the external potentials with $\eta$=0.01 takes more than 20~days of continuous calculations.

\section{Globular clusters tracking a collision on a cosmological timescale}\label{sec:gc-timescale}

\subsection{Globular cluster close passages rate}\label{subsec:cp}

The first step is estimating the GC close passage rate. We adopted two main criteria to characterise the close passages between the GCs. At the same time, (i) the distance between a pair of GCs $dR$ should be less than $100$~pc, and (ii) the relative velocity between a pair of GCs $dV$ should be less than $250$~km~s$^{-1}$. 

In Table \ref{tab:5_tng_ter4} we present the total number of close passages (column two), averaged over all individual 1000 randomisations and for all our different TNG-TVP. We call these events multiple close passages (i.e. an event occurs  a few times during one run between the same two GC1 and GC2). The total number of close passages is quite similar (approximately $\pm$11\%). The minimum number of close passages is obtained for {\tt 441327,} and the maximum events are recorded for the  {\tt 474170} potential. We also defined the average number of individual unique close passage pairs (without multiplicity) for each of our potentials: {\tt 411321} = 4.15, {\tt 441327} = 4.33, {\tt 451323} = 4.18, {\tt 462077} = 4.54, and {\tt 474170} = 3.97. 

\begin{figure}[htbp!]
\centering
\includegraphics[width=0.98\linewidth]{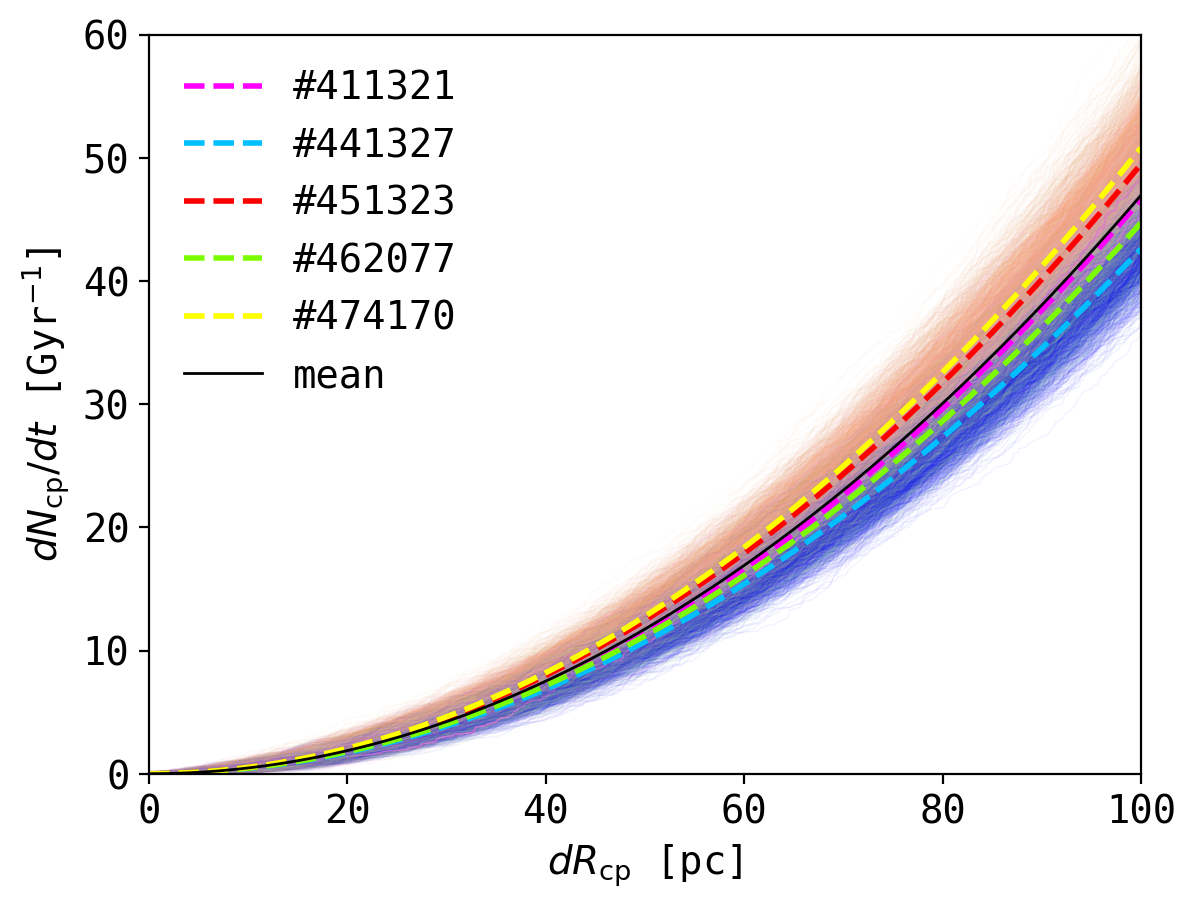}
\caption{Close passages rate of GCs as a function of the relative distance for the 1000 runs in each of the five TNG-TVP (thin solid coloured lines). The dashed lines are the power-law fit (see equation~\ref{eq:fit}) for each TNG-TVP. The solid black line is the mean fitting curve (see Table~\ref{tab:gc-fit}) for all GC orbital close passages.}
\label{fig:statR-GC}
\end{figure}

We marked the GC close passages with cp. Fig.~\ref{fig:statR-GC} shows the cumulative passage number rate as a function ofthe minimum relative distance $dR_{\rm cp}$ of the GCs at the moment of closest approach. The resulting distribution as a function of $dR_{\rm cp}$ can be well fitted by a simple power-law function,
\begin{equation}
\frac{dN_{\rm cp}(dR_{\rm cp})}{dt} = {\rm b}\cdot (dR_{\rm cp})^{\rm a},
\label{eq:fit}
\end{equation}
where the best-fit slope parameters {\bf a} and {\bf b} are compiled in Table~\ref{tab:gc-fit}.

\begin{table}[htbp]
\caption{ Fitting parameters for the GC close passages rate as a function of the relative distance for five TNG-TVP potentials.}
\centering
\begin{tabular}{lll}
\hline
\hline
\multicolumn{1}{c}{Potential} & \multicolumn{1}{c}{\bf a} & \multicolumn{1}{c}{\bf b}  \\
\hline
\hline
{\tt 411321}         & 2.027\;$\pm$\;0.128 & 0.004\;$\pm$\;0.003\\
{\tt 441327}         & 1.991\;$\pm$\;0.112 & 0.004\;$\pm$\;0.003 \\
{\tt 451323}         & 1.990\;$\pm$\;0.107 & 0.005\;$\pm$\;0.003 \\
{\tt 462077}         & 1.998\;$\pm$\;0.109 & 0.005\;$\pm$\;0.003 \\
{\tt 474170}         & 1.990\;$\pm$\;0.104 & 0.005\;$\pm$\;0.003 \\
Mean                   & 1.999\;$\pm$\;0.112 & 0.005\;$\pm$\;0.003 \\
\hline
\end{tabular}
\label{tab:gc-fit}
\end{table}

The rate of the close passages can be described by a simple quadratic relation $dN_{\rm cp}/dt \sim (dR_{\rm cp})^{2}$, which likely is a direct consequence of the overlap of the simple cross-section expression for each GC. Table~\ref{tab:gc-fit} shows that the {\bf a} and {\bf b} slope parameters of the simple power function (equation \ref{eq:fit}) have quite small variances: {\bf a} = 1.999$\pm$0.112 and {\bf b} = 0.005$\pm$0.003, averaged for five TNG-TVPs. By analysing Fig.~\ref{fig:statR-GC}, for example, we can conclude that at a relative distance between the pairs of 50~pc, we can expect about 10 close GCs passages per 1~Gyr on average.

\subsection{Deep passage rates of globular clusters}\label{subsec:deep}

As the next step, we carried out the more detailed analysis of the GC close passages with each other. To do this, we applied the additional condition that the distance between the individual GCs should be less than four times the sums of their half-mass radii: $dR < 4(R_{{\rm hm},i}+R_{{\rm hm},j})$. We call these events collisions. To statistically verify the obtained results reliably, we re-analysed all possible pairs in all 1000 randomisations for five external TNG-TVPs. In Table \ref{tab:5_tng_ter4} we present the total number of collisions (column tree) averaged over all individual 1000 randomisations for all of our different TNG-TVP. The total average numbers of the collisions are quite similar. The largest number of collisions was obtained for {\tt 474170}, and the fewest were obtained for {\tt 441327} potentials. We also defined the number of individual unique collision pairs (without multiplicity) for each run: {\tt 411321} = 2.91; {\tt 441327} = 3.22; {\tt 451323} = 2.97; {\tt 462077} = 3.21 and {\tt 474170} = 3.01.

\begin{table}[htbp]
\caption{Total average number of close passages and collisions in five TNG-TVP potentials (averaged over all individual 1000 randomisations).}
\centering
\begin{tabular}{lccc}
\hline
\hline
\multicolumn{1}{c}{Potential} & {Close passages} & {Collisions} & {Terzan 4 coll.}  \\
\multicolumn{1}{c}{(1)} & {(2)} & {(3)} & {(4)} \\
\hline
\hline
{\tt 411321}         & 462.03 & 43.53 & 5.71 \\
{\tt 441327}         & 427.83 & 40.75 & 4.86 \\
{\tt 451323}         & 496.70 & 47.00 & 6.16 \\
{\tt 462077}         & 449.07 & 41.75 & 5.29 \\
{\tt 474170}         & 510.44 & 48.28 & 5.85 \\
Mean                   & 475.49 & 44.35 & 5.65 \\
\hline
\end{tabular}
\tablefoot{
Column~(1): Names of five TNG-TVPs. 
Column~(2): Total number of events according to criteria (i) and (ii).
Column (3): Collision of events according to criteria (ii) and (iii).
Column (4): Collision of events for the GC Terzan 4 according to criteria  (ii) and (iii).}
\label{tab:5_tng_ter4}
\end{table}

\begin{figure*}[htbp!]
\centering
\includegraphics[width=0.9\linewidth]{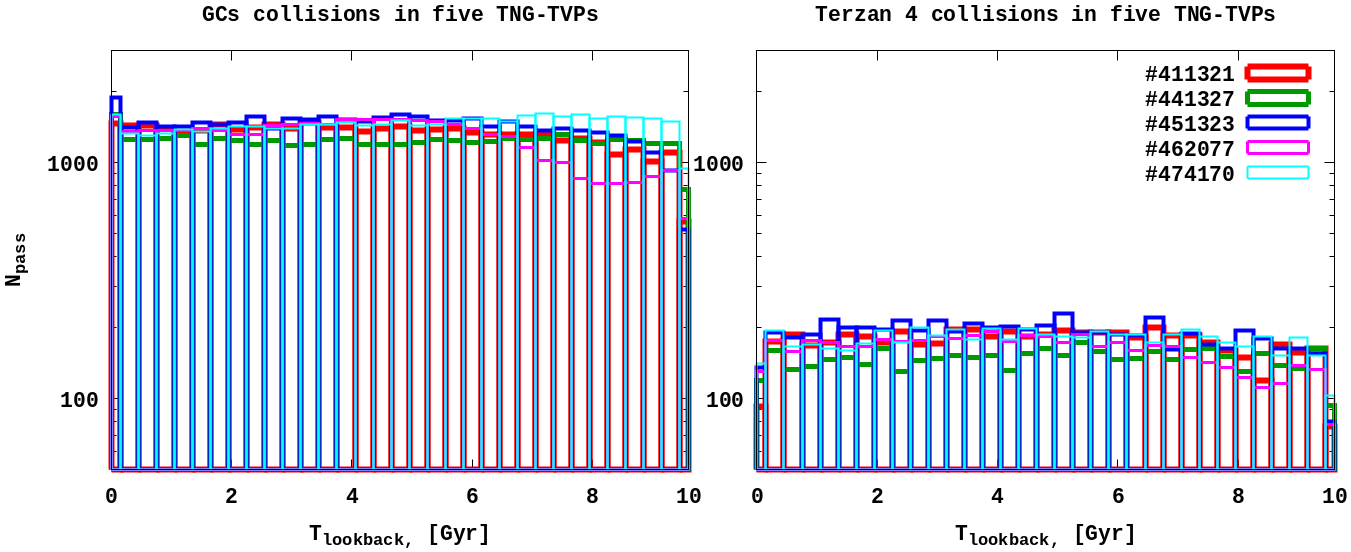}
\caption{Distribution of all possible GC collisions taking the selection criteria (ii) and (iii) during the whole 10 Gyr time interval into account (\textit{left panel}). Same as in the \textit{left panel}, but encounters only with the GC Terzan 4 (\textit{left panel}). The time bin is 0.3 Gyr.}
\label{fig:time-gc-col}
\end{figure*}

In Fig. \ref{fig:time-gc-col} (\textit{left panel}) we present the time distribution for all possible GC collisions, applying the three selection criteria for the full 10 Gyr time interval. Different colours represent each TNG-TVP for each 1000 randomized data sets. The number of total GC events does not depend on the GC global orbital evolution and remains approximately at the same level during the whole backward-integration time. In addition, we can also conclude that the GC collision events for different TNG-TVP potentials are not significantly different.

After the analysis of all 5000 datafiles with the full GC collisions according to our three criteria, we present only pairs with mean probability values highert than 15\% over the five TNG-TVPs. Thus, in Table~\ref{tab:coll-prob-gc} we present the sample of 22~GCs with these unique collision pairs from all five TNG-TVPs. For possible multiple collisions (e.g., five times in the same run between Terzan 4 and NGC 6624), we count these collisions as only one event. In this sense, we define the term ``unique collision'' event.

\begin{table*}[htbp]
\caption{Probability of unique collisions between GCs in five TNG-TVP external potentials applying criteria (ii) and (iii).}
\centering
\sisetup{separate-uncertainty}
\begin{tabular}{
c@{\hspace{0.8\tabcolsep}}
c@{\hspace{0.8\tabcolsep}}
c@{\hspace{0.8\tabcolsep}}
c@{\hspace{0.8\tabcolsep}}
S[table-format=2.1]@{\hspace{0.8\tabcolsep}}
S[table-format=2.1]@{\hspace{0.8\tabcolsep}}
S[table-format=2.1]@{\hspace{0.8\tabcolsep}}
S[table-format=2.1]@{\hspace{0.8\tabcolsep}}
S[table-format=2.1]@{\hspace{0.8\tabcolsep}}
S[table-format=2.1(3)]}
\hline
\hline
GC~1 & $R_{\rm hm1}$ & GC~2 & $R_{\rm hm2}$ & \multicolumn{1}{c}{{\tt 411321}} &  \multicolumn{1}{c}{{\tt 441327}} & \multicolumn{1}{c}{{\tt 451323}} & \multicolumn{1}{c}{{\tt 462077}} & \multicolumn{1}{c}{{\tt 474170}} & \multicolumn{1}{c}{Mean} \\
 & pc & & pc & \multicolumn{1}{c}{\%} & \multicolumn{1}{c}{\%} & \multicolumn{1}{c}{\%} & \multicolumn{1}{c}{\%} & \multicolumn{1}{c}{\%} & \multicolumn{1}{c}{\%} \\
(1) & (2) & (3) & (4) & \multicolumn{1}{c}{(5)} & \multicolumn{1}{c}{(6)} & \multicolumn{1}{c}{(7)} & \multicolumn{1}{c}{(8)} & \multicolumn{1}{c}{(9)} &  \multicolumn{1}{c}{(10)} \\
\hline
\hline
Terzan~4 & 6.06   & Terzan~2   & 4.16 & 51.3 & 43.2 & 56.5 & 47.7 & 46.3 & 49.0\pm5.1  \\
Terzan~4 & \ldots & NGC~6624   & 3.69 & 55.6 & 35.9 & 41.1 & 41.3 & 48.0 & 44.4\pm7.6  \\
Terzan~4 & \ldots & Terzan~5   & 3.77 & 44.6 & 28.7 & 47.9 & 38.8 & 43.7 & 40.7\pm7.5  \\
Terzan~4 & \ldots & NGC~6440   & 2.14 & 35.4 & 33.3 & 45.0 & 41.7 & 47.7 & 40.6\pm6.2  \\
Terzan~4 & \ldots & Liller~1   & 2.01 & 56.3 & 33.1 & 45.7 & 38.0 & 39.6 & 42.5\pm8.9  \\
Terzan~4 & \ldots & NGC~6528   & 2.73 & 26.2 & 16.5 & 31.9 & 18.1 & 21.1 & 22.8\pm6.3  \\
Terzan~4 & \ldots & NGC~6558   & 1.70 & 13.3 & 19.7 & 18.3 & 17.8 & 22.0 & 18.2\pm3.2  \\
Terzan~4 & \ldots & NGC~6638   & 3.69 & 14.2 & 20.8 & 11.0 & 15.2 & 17.9 & 15.8\pm3.7  \\
Terzan~4 & \ldots & NGC~6642   & 1.51 & 10.8 & 20.8 & 13.9 & 18.5 & 14.2 & 15.6\pm4.0  \\

Terzan~2 &  4.16  &  Djorg~2   & 5.16 & 19.6 & 19.2 & 42.7 & 30.1 & 37.3 & 29.8\pm10.5  \\
Terzan~2 & \ldots & Liller~1   & 2.01 & 19.8 & 17.6 & 40.2 & 29.6 & 37.8 & 29.0\pm10.2  \\
Terzan~2 & \ldots & Terzan~5   & 3.77 & 26.4 & 19.9 & 19.2 & 20.0 & 22.2 & 21.5\pm2.9  \\ 
Terzan~2 & \ldots & NGC~6440   & 2.14 & 16.5 & 15.9 & 26.7 & 20.1 & 25.4 & 20.9\pm5.0  \\
Terzan~2 & \ldots & Terzan~6   & 1.33 & 14.0 & 14.9 & 25.5 & 16.5 & 19.9 & 18.2\pm4.7  \\

Terzan~5 & \ldots & NGC~6540   & 5.32 & 13.2 & 11.6 & 20.3 & 17.2 & 19.1 & 16.3\pm3.8  \\ 

Liller~1 & 2.01   &  Djorg~2   & 5.16 & 10.4 & 11.4 & 23.6 & 16.1 & 20.8 & 16.5\pm5.8  \\
Liller~1 & \ldots & Terzan~5   & 3.77 & 14.8 & 17.3 & 20.2 & 18.5 & 19.6 & 18.1\pm2.1  \\

NGC~6624 & 3.69   & HP~1       & 6.06 & 10.3 &  8.0 & 23.6 & 21.4 & 16.5 & 16.0\pm6.8  \\
NGC~6218 & 4.05   & NGC~6254   & 4.81 &  4.7 & 22.4 & 30.1 &  8.6 & 24.8 & 18.1\pm10.9  \\ 
NGC~6256 & 4.82   & NGC~6540   & 5.32 & 22.4 & 11.3 & 23.8 & 17.1 & 25.1 & 19.9\pm5.7  \\
NGC~6256 & \ldots & NGC~6304   & 4.26 &  9.9 & 11.1 & 22.1 & 15.2 & 18.2 & 15.3\pm5.0  \\
NGC~6540 & 5.32   & NGC~6717   & 4.23 & 37.0 & 10.0 &  7.4 &  9.9 & 11.0 & 15.1\pm12.3  \\
\hline
\end{tabular}
\tablefoot{
Columns~(1) and (3) represent the names of the collision GCs, and columns~(2) and~(4) list their half-mass radii. Columns~(5)-(9) list the collision probabilities for GCs pairs in all five TNG-TVPs in percent. Column~(10) contains the average probability value with the error over all potentials.}
\label{tab:coll-prob-gc}
\end{table*} 

Table~\ref{tab:coll-prob-gc} shows that statistically, the obtained collision probability is noticeable and can reach for individual pairs a mean level of $\approx40\%$ for all the five TNG-TVPs, even considering that we used the present-day half-mass radii of hte GCs. Taking into account that in the past, GCs had larger half-mass radii \citep[][see Figure 2]{MHS2012}, we can assume that GCs can have an even higher probability of collisions in the past.  Table~\ref{tab:coll-prob-gc} shows that the most active GCs (with a noticeable collision probability) from our sample are Terzan~4, Terzan~2, NGC~6624, Liller~1, and NGC~6624. As an example, we present in Table \ref{tab:5_tng_ter4} the total number of events with Terzan 4 in all five TNG external potentials. Fig. \ref{fig:time-gc-col} (\textit{right panel}) shows the time distribution for the GC Terzan 4 collisions, applying our three criteria during the whole 10 Gyr interval, as for the most active GC. We also conclude that there is no significant difference between the GC collision events for different TNG-TVP potentials.

We additionally estimated the impact of the random sample sizes on the results obtained in Table~\ref{tab:coll-prob-gc}. For this investigation, we randomly sampled the total 1000 random realisations for  200, 400, 600, 800, and 1000  in {\tt 411321} TNG-TVP potential. As expected, starting from 200 sample size, the collision results practically no longer depend on the sample size. This means that starting from a few hundred realisations, the collision numbers are saturated. As an example, we show the Terzan~4 versus Terzan~2 collision probabilities: 59\% (sample size 200), 49.75\% (400), 62\% (600), 58.13\% (800), and 51.3\% (1000). The Terzan 4 versus NGC 6624 collision probabilities are 40.5\% (sample size 200), 41.25\% (400), 55\% (600), 51.62\% (800), and 55.6\% (1000).

\begin{figure*}[htbp!]
\centering
\includegraphics[width=0.9\linewidth]{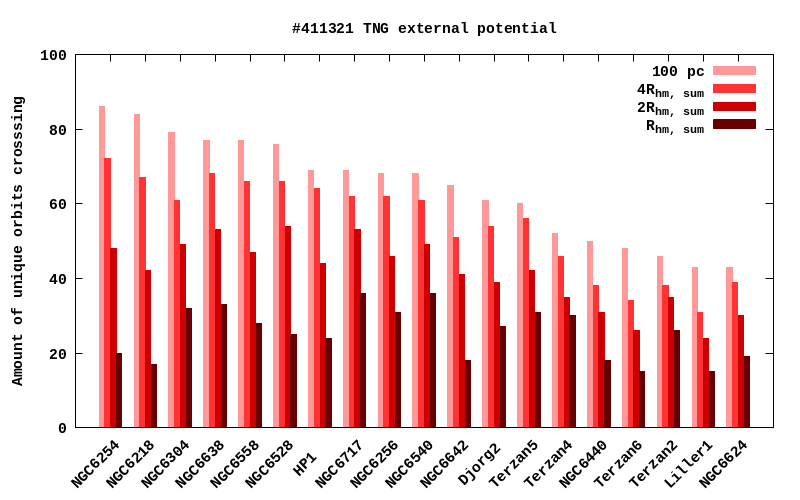}
\caption{Number of unique orbit crossings for selected clusters from Table \ref{tab:coll-prob-gc} for four different distance criteria and for the selected {\tt 411321} TNG-TVP external potential.}
\label{fig:gc-col}
\end{figure*}

We estimated the number of unique close passages and collisions for each individual cluster listed in Table \ref{tab:coll-prob-gc}, using four different distance criteria: $N_{100\rm pc}$, $N_{4R_{\rm sum}}$, $N_{2R_{\rm sum}}$ and $N_{R_{\rm sum}}$, see Fig.~\ref{fig:gc-col}. 
We counted all the close encounters and collision events for the 19 selected GCs with all the other GCs, including all the pairs even with a low probability (lower then 15\%). In this figure, we obtain different numbers for each of these criteria. As an illustration, we provide this additional  analysis for the {\tt 411321} TNG-TVP external potential.

 Fig.~\ref{fig:gc-col} shows that the different criteria give us the different unique crossing (collision) numbers. They change proportionally, however, depending on the relative distance criteria. The highest value of a unique crossing number is reached for the full 100 pc distance criteria. The $4~R_{\rm sum}$ also gives us comparable numbers, however. The $2~R_{\rm sum}$ and $R_{\rm sum} = R_{{\rm hm},i}+R_{{\rm hm},j}$ criteria proportionally give us smaller numbers. The changes are smooth, which means that we properly resolved all the unique orbit crossings in our full dynamical MW GC system. 

The actual collision percentage probability between GCs here are quite different from our earlier results based on the initial data of the \textit{Gaia} DR2 catalogue. 
We can easily compare our current results with our previous work \citep{Ishchenko2023-kfnt} because we previously used the same conditions of integration, external potentials, and scheme of collision analysis. For example, in Table~\ref{tab:coll-prob-gc} for the pair Terzan 4 versus NGC 6440, we now obtain a mean value of 40\%, and according to the DR2, we obtain 26\%; for Terzan 4 versus Terzan 5, we now obtain 40\% in contrast to 18\%; and for the Terzan 2 versus Terzan 4 pair, we obtain 49\% compared with GR2, 22\%. More example comparisons can be found in Table 3 in \cite{Ishchenko2023-kfnt}. Thus, the increase in the accuracy of the observed positions, and especially of the proper motions and velocities of the clusters, has increased the GC collision events by almost a factor of two. This significant increase in the collision event numbers is probably a pure chance effect.

In Appendix \ref{app:ter4_orb} we present the Terzan 4 orbital evolution with the collision points with other GCs in the {\tt 411321} TNG external potential, as presented in  Table~\ref{tab:coll-prob-gc}. Collision events for Terzan 4 with a statistical probability higher than 25\% in this potential are shown by colour-coded circles. 

\subsection{Distribution of the globular cluster probability}\label{subsec:prob}

\begin{figure*}[htbp!]
\centering
\includegraphics[width=0.99\linewidth]{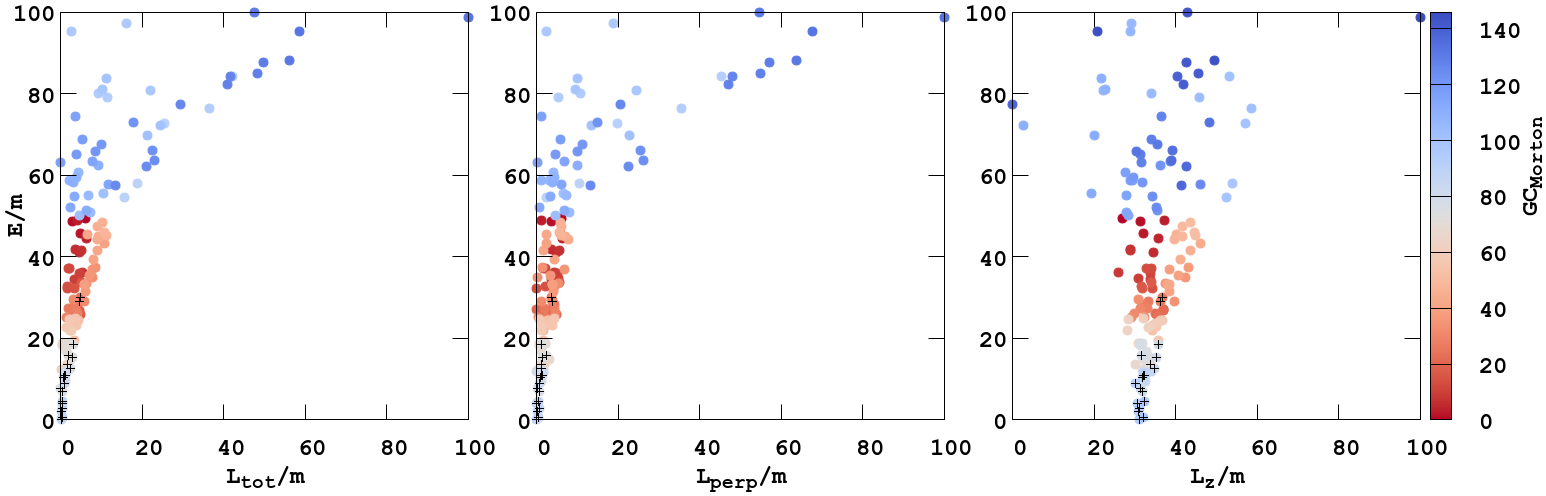}
\caption{Projections of the individual GCs positions from the 3D phase space of the normalised specific values of ($E$, $L_{tot}$, $L_{perp}$, and $L_{z}$) to ($L_{tot}$ and $E$) plane in the \textit{left} and \textit{centre} panel ($L_{perp}$, $E$), and in the ($L_{z}$, $E$) plane in the \textit{right} panel. The data are represented for the {\tt 411321} TNG-TVP potential after sorting by the Morton space-filling curve, where the colour indicates the order of the GCs after the Morton analysis. The black crosses indicate the selected GCs from Fig.~\ref{fig:gc-col}.}
\label{fig:pair-gc-3d-pot}
\end{figure*}

\begin{figure*}[htbp!]
\centering
\includegraphics[width=0.99\linewidth]{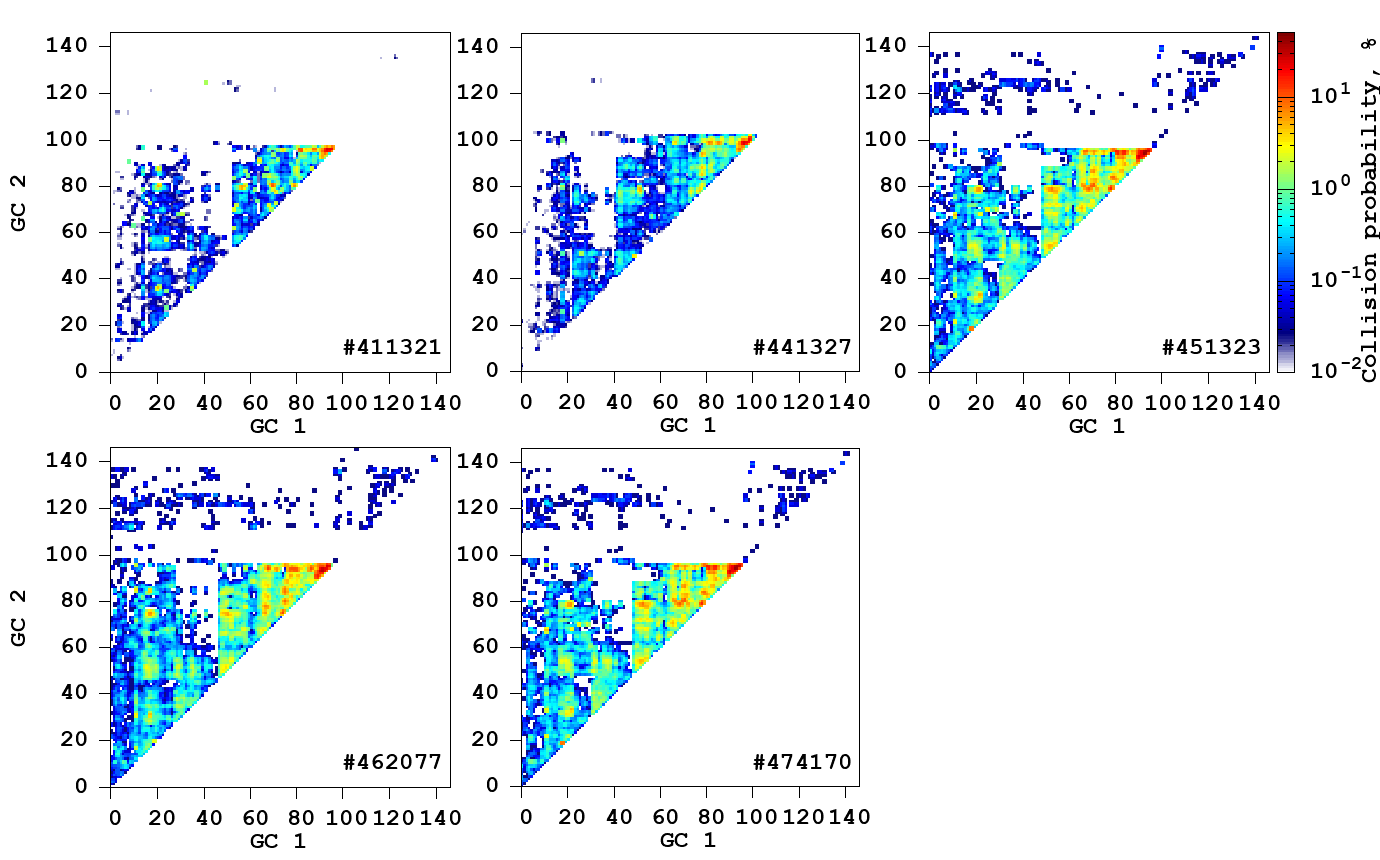}
\caption{ Statistical probabilities of GC collisions with each other for five external TNG-TVP potentials after sorting by Morton space-filling curve. As input data for the Morton space sorting, we used the values of total specific energy $E$, total specific angular momentum $L,$ and the $z$th component of the specific angular momentum $L_{z}$ for each GC. The colour indicates the percent of the individual GC collision probability. The $X$ - and $Y$ -axes show the new GC indices obtained after the Morton space ordering.}
\label{fig:pair-gc-3d}
\end{figure*}

Our third step was to perform a visual analysis of the statistical probability of the GC collisions in the current (at the present time) specific energy and specific angular momentum 3D phase space ($E$, $L_{tot}$, $L_{z}$) of the GCs. To sort the data (clustering), we used a 3D Morton space-filling curve \citep{Lebesgue1904, Morton1966, Sagan1994}. Using these phase-space values after normalisation (i.e. fitting the specific energy and specific angular momentum ranges from 0 to 100), we sorted all our GCs into this constructed one-dimensional distribution (i.e. we obtained the Morton sorted order (number) GC$_{\rm Morton}$ for each GC). 
In Fig.~\ref{fig:pair-gc-3d-pot} we present the GC distribution in our 
3D phase space, colour-coded by appropriate GC$_{\rm Morton}$ sorted number for the potential {\tt 411321}. We also show the GCs with black crosses (from Fig.~\ref{fig:gc-col}) that show a close collisions. These GCs are well clustered in the middle of our Morton ordered list (around low values of $E$ and $L_{tot}$). In Fig.~\ref{fig:pair-gc-3d} we present the analysis of statistical probability of collisions, based on the values of GC$_{\rm Morton}$ sorted numbers GC1 and GC2 for all five TNG-TVP. We excluded the case of GC1 $\equiv$ GC2 from the collision count. 
We show the statistical probability of GC collisions in percent according to the adopted criteria (ii) and (iii). 
The probability distribution in Fig.~\ref{fig:pair-gc-3d} shows that certain groups of GCs, sorted by physical parameters (energy, total angular momentum, and $z$th component of the angular momentum) form a fairly high percentage of the GCs collisions, more than 30\%.
Based on these plots, we can conclude that the collisions with the highest probability in all five potentials are are concentrated between the GC$_{\rm Morton}$ sorted numbers 60 to 100. These collisions are almost all listed in our Table~\ref{tab:coll-prob-gc} and are also presented in the Fig.~\ref{fig:gc-col}.

\begin{figure*}[htbp!]
\centering
\includegraphics[width=0.49\linewidth]{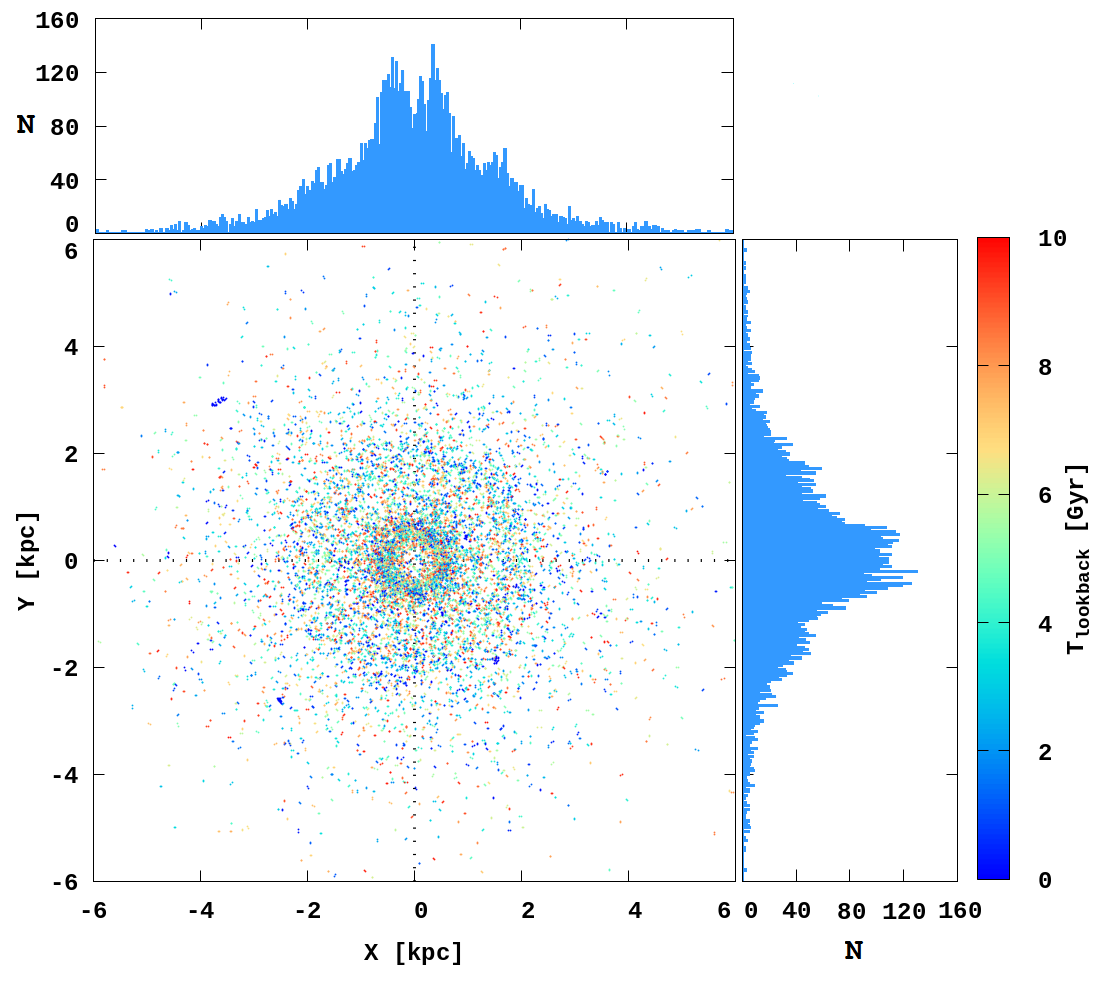}
\includegraphics[width=0.49\linewidth]{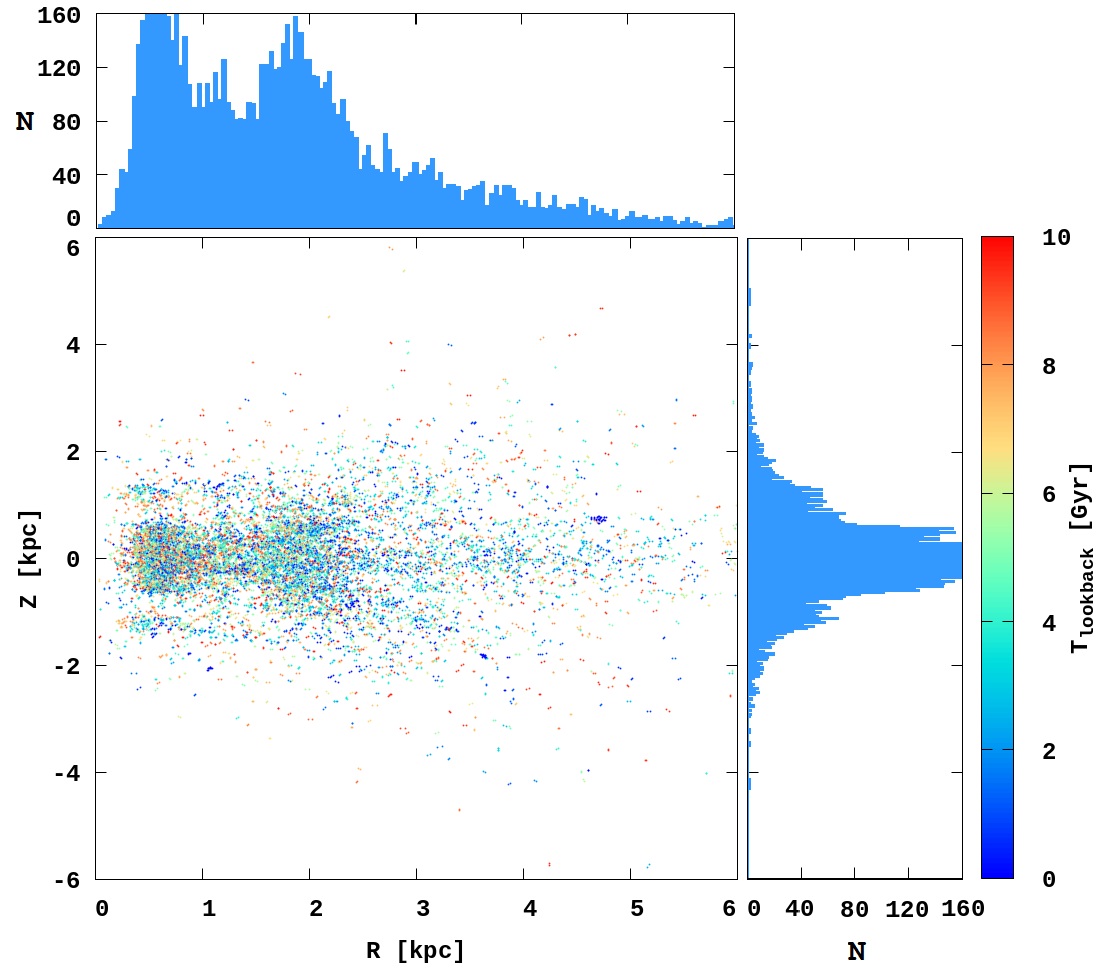}
\caption{Distribution of GC collision pairs for {\tt 411321} TNG-TVP external potential. The colour of the points represents the lookback time of these events. The histogram presents the projected number density of the all collision events (criteria (ii) and (iii)) independently of the lookback time.}
\label{fig:gc-col-orb}
\end{figure*}

We also present the distribution of collision points in Fig.~\ref{fig:gc-col-orb} in a 3D Cartesian galactocentric coordinate system. Each point represents the individual collision events (criteria (ii) and (iii)). The colours of the points represent the lookback time of these events. 

Most collisions are located inside the Galactic disk at a cylindrical distance shorter than $\sim4$~kpc. The distribution of the collision point distance from the Galactic plane is also well below $\pm2$~kpc. The ($X$, $Y$) distribution shows that the collision points are located in a ring-like structure. The first ring-like structure has the highest collision number density at $\sim1$~kpc, and the second ring structure has a maximum at $\sim2$~kpc.
These rings are most probably connected with the pericentre and apocentre values of our most active collisional GCs (e.g. Terzan 4, NGC 6624, and Terzan 2; see the webpage of the project \footnote{Orbits for all 159 GCs in TNG-TVP potentials: \url{https://sites.google.com/view/mw-type-sub-halos-from-illustr/GC-TNG}.}). Several dozen GC collision lie very close to the Galactic centre (see more details in our previous publication \cite{Ishchenko2023c}).

\section{Conclusions}\label{sec:con}

The evolution of our Galaxy GC subsystem is one of the most discussed topics of modern day astrophysics: they help us to understand the early Galaxy, the history of the Galaxy formation due to accretion events, the evolution itself, and and so on. Our study of the GC time-evolved orbital evolution is an attempt to proceed in this field of investigation. In this case, in the third part of the series of our papers, we presented the analysis of the statistical probability of GC close passages and even collisions with each other. 

To perform our GC orbital integration in the realistic model of our Galaxy, we used the mass distribution data from the IllustrisTNG-100 cosmological modelling database. We selected five Milky Way-like time-variable external potentials that have the closest parameters to our Galaxy at the present time: {\tt 411321, 441327, 451323, 462077, and 474170}.  
To estimate the probability of individual GC close orbital passages with each other during their evolution, we created a set of 1000 initial randomised conditions. We varied the three velocity components (on the base of the errors in the proper motions and radial velocity) to obtain the velocity in the Cartesian galactocentric rest-frame coordinate system for each individual GC. This approach yielded reliable results.  

For the individual GC orbital integration, we used the high-order parallel dynamical $N$-body code $\varphi$-GPU, which is based on the fourth-order Hermite integration scheme with hierarchical individual block time steps. Each GC was integrated as a separate $N$-body particle with its own mass (taking the GC versus GC gravitational interaction into account as well). The integration of all the 147 GC orbits was performed up to 10~Gyr lookback time.

As a definition of the close passages between the GCs, we adopted two criteria. At the same time, (i) the distance between the GCs $dR_{\rm cp}$ should be smaller than 100~pc, and (ii) the relative velocity between these objects $dV_{\rm cp}$ should be lower than 250~km~s$^{-1}$.

As a definition of the collision event between the GCs, we added one more criterion: At the same time condition (ii), the separation between the two GCs falls below four times the sums of their half-mass radii (iii).
The main results of this study are listed below.

\begin{itemize}

    \item Applying criteria (i) and (ii), we can conclude that for example at the relative distance of 50 pc between GCs, we can expect about 10 close passages during every 1 Gyr for each of the five MW-like TNG-TVP external potentials on average. At a distance of 80 pc, we can expect already above 30 close passages for the same time interval. The cumulative close passage rate can be very well approximated by the simple quadratic power law, which clearly reflects the quadratic cross-section nature of the GC versus GC possible close interactions. 

    \item We defined the average number of individual unique close passages pairs (without multiplicity) for each runs: {\tt 411321} = 4.15, {\tt 441327} = 4.33, {\tt 451323} = 4.18, {\tt 462077} = 4.54, and {\tt 474170} = 3.97.

    \item Applying criterion (iii), we estimated the statistical probability of these events and found the unique collision (without multiplicity) number for each TNG-TVP: {\tt 411321} -- 2.91, {\tt 441327} -- 3.22, {\tt 451323} -- 2.97, {\tt 462077} -- 3.21, and {\tt 474170} -- 3.01.

    \item We estimated the statistical probability of these events and found the 22 most reliable collision pairs. The most noticeable probability pairs (in percent) were obtained for Terzan~4 versus Terzan 2 (49\%), Terzan~4 versus NGC~6624 (44\%), Terzan~4 versus Terzan~5 (40\%), Terzan~4 versus NGC 6440 (40\%), and Terzan~4 versus Liller~1 (42\%). In the  sense of collisions, the most active GC in our sample is Terzan 4, which has 5.65 collision events for all five TNG-TVPs on average.  

    \item All our selected GCs that undergo a collision event belong to the thick disk of our MW (up to $\sim$1~kpc in $z$ direction and $\sim$2~kpc in $R$), according to \cite{BHG2016}.

    \item In many cases, the collisions between GCs occur at separations that are significantly smaller than four sums of their half-mass radii, and sometimes even less. For example, we have 58 individual cross sections for a separation limit of 100~pc and 38 pc for GC Terzan 4 when we simply even sum their half-mass radii. Also as an example, the GC Terzan 2 has 39 individual cross sections for a separation limit of 100~pc and 30 pc when we simply sum their half-mass radii. 

    \item To visually present the  statistical probability analysis of the GC collisions, we used a 3D Morton space-filling curve to sort the GCs based on their values of total energy $E$, total angular momentum $L_{tot}$ , and the $z$th component of the angular momentum $L_{z}$. Based on our visual analysis (of the sorted GCs), we find separate groups of GCs with a fairly high percentage of GC collisions. The probability is higher than 30\% over all random realisations.

    \item We also presented the distribution of collision points in a 3D Cartesian galactocentric coordinate system. Most of the collisions are located inside the Galactic disk at a cylindrical distance smaller than $\sim4$~kpc. The distribution of the collision point distance from the Galactic plane is also well smaller than $\pm2$~kpc. The collision points form two ring-like structures. The first ring-like structure has the highest collision number density at $\sim1$~kpc, and the second ring structure has a maximum at $\sim2$~kpc.

\end{itemize}

Based on our current simple calculations, we conclude that the GC versus GC collisions in our Milky~Way are relatively common events. In quite different TNG-TVP Milky~Way-like external potentials, the frequency of the collision events is similar. The implication of our results would be interesting for the investigation of Galaxy global GC dynamical evolution on a cosmological timescale.    

\begin{acknowledgements}
The authors thank the anonymous referee for a very constructive report and suggestions that helped significantly improve the quality of the manuscript.

The authors are thanks to Sergey Khoperskov for a valuable discussions during the whole period of work on the paper. 

The MI, PB, CO, MK and DY acknowledge the support within the grant No.~AP14869395 of the Science Committee of the Ministry of Science and Higher Education of Kazakhstan (``Triune model of Galactic center dynamical evolution on cosmological time scale''). 

The work of MI, MS, PB and OS was supported under the special program of the NRF of Ukraine ``Leading and Young Scientists Research Support'' -- ``Astrophysical Relativistic Galactic Objects (ARGO): life cycle of active nucleus'', No. 2020.02/0346.

The work of MI was supported by the Grant of the National Academy of Sciences of Ukraine for young scientists.

The work of PB was supported by the Volkswagen Foundation under the special stipend No.~9D154. 

MS acknowledges the support under the Fellowship of the President of Ukraine for young scientists 2022-2024.

The work of PB, MI, MS and OS was supported under the special program ``Long-term program of support of the Ukrainian research teams at the PAS Polish Academy of Sciences carried out in collaboration with the U.S. National Academy of Sciences with the financial support of external partners''. 

This work has made use of data from the European Space Agency (ESA) mission GAIA (\url{https://www.cosmos.esa.int/gaia}), processed by the GAIA Data Processing and Analysis Consortium (DPAC, \url{https://www.cosmos.esa.int/web/gaia/dpac/consortium}). Funding for the DPAC has been provided by national institutions, in particular the institutions participating in the GAIA Multilateral Agreement. 
\end{acknowledgements}

\bibliographystyle{mnras}  
\bibliography{part_3}   

\begin{appendix}

\section{Orbital evolution of the globular cluster Terzan 4 with the collision moments.}\label{app:ter4_orb}

As an example, we present the orbital evolution of the most active GC (in the collisional sense) Tezan 4 with its GC counterparts during the whole 10 Gyr time interval. 

\begin{figure*}[htbp]
\centering
\includegraphics[width=0.95\linewidth]{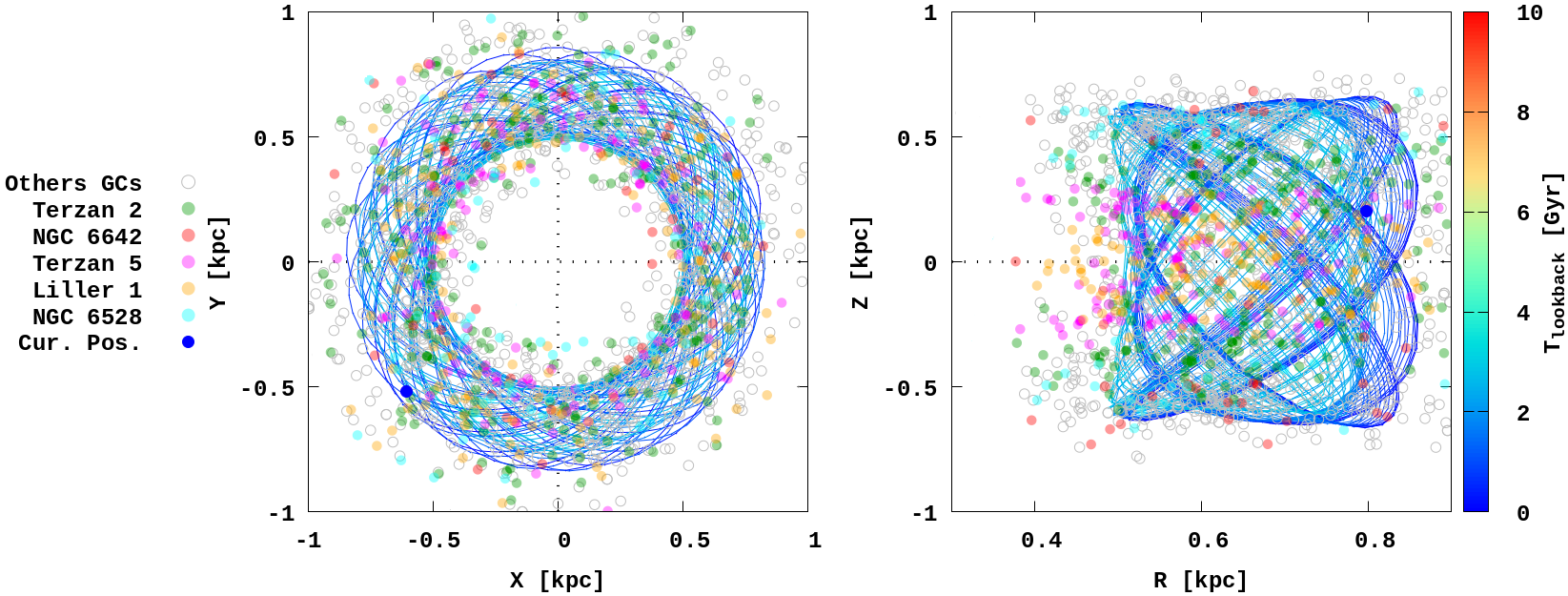}
\includegraphics[width=0.95\linewidth]{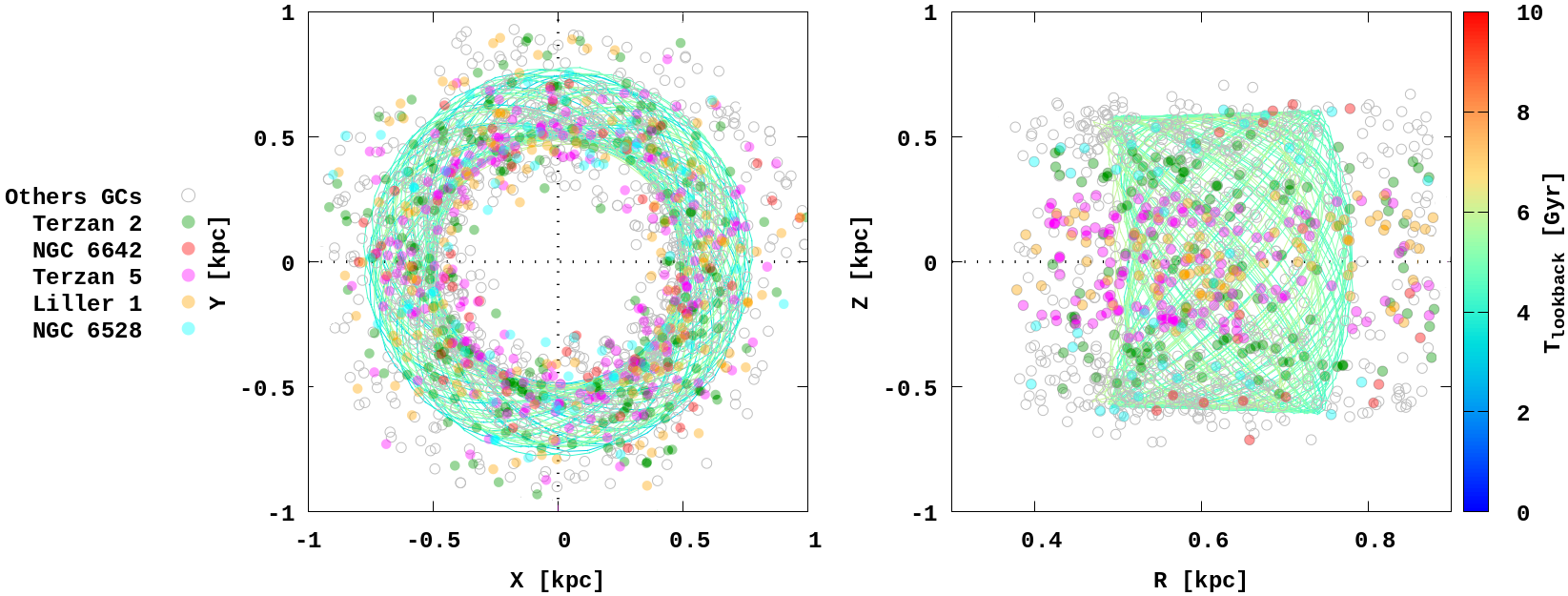}
\includegraphics[width=0.95\linewidth]{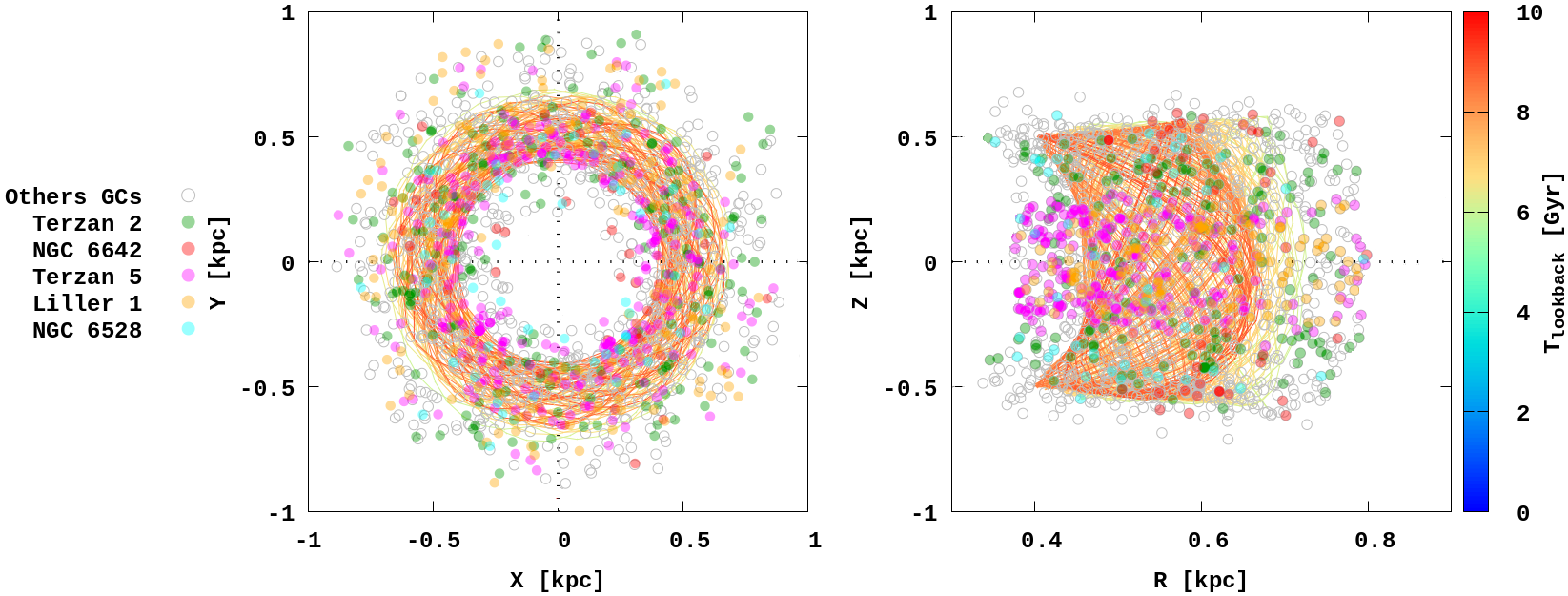}
\caption{Pairs for Terzan 4 with a statistical probability higher than 25\% in {\tt 411321} TNG-TVP. Green circles whos Terzan 2, red circles show NGC 6642, magenta circles show Terzan 5, orange circles show Liller 1, and cyan circles whos NGC 6528. Empty grey circles show the other GCs. The blue points are the start and end of the integration. \textit{From top to bottom, the panels}  represent the interval of integration: 0 -- 3 Gyr, 3 -- 6 Gyr, and 6 -- 9 Gyr in lookback time.}
\label{fig:ter4_coll1}
\end{figure*}

\end{appendix}

\end{document}